\documentclass[journal,comsoc]{IEEEtran}
\usepackage[T1]{fontenc}

\ifCLASSINFOpdf
\else
\fi
\usepackage{amsmath}
\usepackage[cmintegrals]{newtxmath}
\usepackage{graphicx}
\usepackage{caption}
\usepackage{subcaption}
\usepackage{psfrag,booktabs}
\usepackage{float}
\usepackage{multirow}
\usepackage[table]{xcolor}
\usepackage{array}
\usepackage{filecontents}

\begin{document}
\bstctlcite{IEEEexample:BSTcontrol}
\title{Exploiting WiFi Channel State Information for Residential Healthcare Informatics}

\author{
\IEEEauthorblockN{{Bo Tan}\IEEEauthorrefmark{1}\IEEEauthorrefmark{2}}
\IEEEauthorblockN{{Qingchao Chen}\IEEEauthorrefmark{2}}
\IEEEauthorblockN{{Kevin Chetty}\IEEEauthorrefmark{2}}
\IEEEauthorblockN{{Karl Woodbridge}\IEEEauthorrefmark{2}}
\IEEEauthorblockN{{Wenda Li}\IEEEauthorrefmark{3}}
\IEEEauthorblockN{{Robert Piechocki}\IEEEauthorrefmark{3}}
\\
\IEEEauthorblockA{\IEEEauthorrefmark{1}Coventry University}
\IEEEauthorblockA{\IEEEauthorrefmark{2}University College London}
\IEEEauthorblockA{\IEEEauthorrefmark{3}University of Bristol}
}

\maketitle

\begin{abstract}
Detection and interpretation of human activities have emerged as a challenging healthcare problem in areas such as assisted living and remote monitoring. Besides traditional approaches that rely on wearable devices and camera systems, WiFi based technologies are evolving as a promising solution for indoor monitoring and activity recognition. This is, in part, due to the pervasive nature of WiFi in residential settings such as homes and care facilities, and unobtrusive nature of WiFi based sensing. Advanced signal processing techniques can accurately extract WiFi channel status information (CSI) using commercial off-the-shelf (COTS) devices or bespoke hardware. This includes phase variations, frequency shifts and signal levels. In this paper, we describe the healthcare application of Doppler shifts in the WiFi CSI, caused by human activities which take place in the signal coverage area. The technique is shown to recognize different types of human activities and behaviour and be very suitable for applications in healthcare. Three experimental case studies are presented to illustrate the capabilities of WiFi CSI Doppler sensing in assisted living and residential care environments. We also discuss the potential opportunities and practical challenges for real-world scenarios.   
\end{abstract}


\IEEEpeerreviewmaketitle

\section{Introduction} \label{s:introduction}

\IEEEPARstart{H}{ealthcare} demands are becoming an increasing concern in modern society due to the ageing population, and rising levels of obesity, cardiovascular disease, depression and mental health problems. This has led to significant resource and financial constraints on healthcare services globally. New technologies for sensing daily activities and behaviours in residential settings and care homes are providing insightful data relating to both pattern-of-life, and shorter term monitoring such as inactivity and falls. These metrics are especially useful for identifying health issues and chronic diseases, for which early treatment interventions are critical.

In this context, measuring and modelling human behaviour and routine activities with both passive and active in-home sensors has become increasingly important.  The SPHERE Healthcare project \cite{SPHERE_book} has designed and developed a multi-modal sensor system that will be deployed across 100 residential homes in Bristol, UK for monitoring the daily activities of participants. The overall system leverages a wide range of sensors including passive infra-red (PIR) detectors, wearable accelerometers, motion cameras, and WiFi based passive gesture sensors. Among the sensing technologies used in human activity modelling, 802.11 (WiFi) based activity sensing is drawing significant research attention owing to operational advantages brought about by its ubiquitous nature, and strong signal coverage throughout homes and urban environments more generally. Additionally, the unobtrusive nature of the technology, and its inability to generate an image of person is favourable in terms of privacy, which is an important concern of many end-users.
 
The presence of a person (either static or in motion) within a WiFi signal field affects the characteristics of the communication channel, for example by increasing propagation paths, attenuating the signal, inducing a frequency shift etc. This results in time-varying characteristics that corresponding to real-time body movements or physical gestures. Channel state information (CSI) is typically used to describe signal propagation properties which include the distortions induced by human activities. This has stimulated research into the use of CSI data obtain from off-the-shelf network cards or customized systems to interpret various types of human behaviour.

Traditional received signal strength (RSS) based methods have demonstrated the capability of using WiFi signals for localization and activity recognition in residential areas. However, these systems must undergo labour intensive offline training, and suffer from coarse resolution. Researchers have therefore begun to leverage WiFi CSI to obtain more accuracy and a larger application space. This approach differs from wireless and mobile communications which typically employs the statistical characteristics of the wirelese channel, for example Rician, Ryleigh and Nakagami channels \cite{Statistical_channel_model}. Several advances have recently been made in WiFi CSI research. A subspace based method entitled Matrack \cite{Matrack} utilises angle of arrival (AoA) and time of arrival (ToA) measurements for fine-grained device free localisation. In \cite{RT_Fall} the authors employ phase information and RSS profiles to identify people falling for applications in tele-healthcare, while the approach proposed by Zhang \cite{Fresnel_2} first exploits Fresnel zones to examine the effect of the location and orientation in human respiration monitoring and detect human walking direction. CSI measurements have also been integrated into wearable technologies such as Headscan \cite{HeadScan} to monitor activities which involve torso, head and mouth movements, and Bodyscan \cite{BodyScan} for sensing everyday activities, including vital signs. Here, the authors demonstrate activity recognition while subjects carried out activities such as walking, bending over, shaking, chewing, coughing and drinking.

Within the literature concerning CSI based activity recognition, the Doppler information in the WiFi channel state arising from human activities has gained significant attenuation. For example, WifiU \cite{WiFiU} can identify WiFi Doppler shifts for in-home human gait recognition applications by filtering out the high frequency jitters and denoising of WiFi CSI profile. Passive WiFi detection is another Doppler-based methodology and in \cite{Passive_WiFi_TAES} the authors demonstrate the through-wall detection capability by applying cross-ambiguity function processing on WiFi signals reflected from targets of interest. In \cite{CARM}, researchers use the Short-Time Fourier Transform (STFT) and Discrete Wavelet Transform (DWT) to separate reflections from different body parts in the frequency domain. Then, activity is modelled by profiling the energy each frequency component. Considering the parameters inherent to WiFi CSI such as the time delay, RSSI and phase shift/frequency, the Doppler shift or the frequency component is the only metric that reflects dynamic states and thus has the potential to monitor the movements of people and objects within WiFi enabled areas.

In some cases, CSI can be used directly to determine behaviour metrics, for example, respiration rates of personnel \cite{UCL_breath}, or to capture the trace of a moving hand by synthesizing AoA information \cite{Matrack}. In many cases however, parameters within the CSI matrix cannot be easily used to make inferences about the type of activity or human behaviour occurring. In these cases, machine learning methods are applied to analyse the pattern of CSI changes and related these to gesture and other activities. In \cite{BodyScan}, statistical classification methods such as support vector machine (SVM) and k-nearest neighbour (k-NN) are used to recognize key strokes on a keyboard and hand waving gestures. Some studies have also determined a time sequential relation between two or more human gestures or activities. A Hidden Markov Model (HMM) is introduced in \cite{CARM} to improve recognition performance.

This study is concerned with the exploitation of frequency/Doppler data in the WiFi CSI to provide information on human activities within indoor environments, with a focus on healthcare monitoring. The paper is organized as follows: firstly in Section \ref{s:challenges} we outline key challenges that exist with activity and behaviour recognition within residential healthcare. Section \ref{s:dsp_br} then describes how human activities impact the WiFi CSI parameters, and the subsequent methodology for performing the recognition. In Section \ref{a:case_study}, three case studies are provided to illustrate the use of Doppler based WiFi CSI for activity reconition in healthcare monitoring. Finally, in Section \ref{s:conclusion}, we discuss the implications of our results and possible directions for future research before summarizing the main finding of the work.

\section{Challenges of Behavior Recognition in Residential Healthcare} \label{s:challenges}

\subsection{Challenges in Residential Health Monitoring}
\subsubsection{Vital Signs}
Vital signs such as respiration and the heart beat are the amongst the most useful indicators of a person's general health. Equipment like chest belts, electrocardiogram (ECG) or photoplethysmogram (PPG) sensors can accurately record respiration and heart rates and are used in some specific controlled scenarios such as  hospitals and care homes. However, there is a low acceptance level associated with these devices in normal residential applications due to the inconvenience it causes when integrated into everyday routines. Newer wearable devices, like smart watches and wristbands can keep a record a user's heart rate during sport by using a physiological sensor. However, users often use these for a specific purpose e.g. exercising and therefore go through long periods of not being monitored. There is therefore a strong requirement for long term daily monitoring of vital signs. 
\subsubsection{Life Threatening Events}
Besides vital signs, detection of events that could lead to serious injuries, and even fatalities is another important aspect in residential healthcare. Fall detection in one of the most critical capabilities in this context. Currently available fall detection alert devices rely on the user wearing a wristband or necklace and use on-board accelerometer and gyroscope data to indicate sudden changes of movement, direction or position. However, wearable devices tend to have limited battery life and have high risk of non-compliance, especially amongst the elderly. Thus, solutions are needed that can provide seamless monitoring of life threatening events in residential settings especially those which accommodate the elderly or people with mobility disabilities.
\subsubsection{Daily Activities}
A log of daily activities is of great interest in healthcare because it contains diverse health related information. For example, the activity of making a cup of tea will be an indication of water intake, and potentially even sugar and milk. In other approaches, data from different smart home sensors were fused. This included: water, gas, electricity meters and even humidity readings to recognize physical activities. There are other sensors in smart homes which can directly monitor human motion, such as PIR sensors, but these only deliver limited information that indicate presence of a person. Camera or other optical sensors can accurately recognize gestures and activities, but under perform in suboptimal light conditions and raise privacy concerns. Wearable devices like accelerometer and on body RF sensors \cite{BodyScan}, \cite{HeadScan} can be uncomfortable, and are liable to be misplaced, damaged or forgotten. Unobtrusive monitoring technology like WiFi based gesture recognition that can fill these gaps is therefore a promising area to explore.
\subsubsection{Chronic Activity Level}
Apart from the above information, longer term activity data over days, weeks even months, is also valuable in the context of healthcare. Research in \cite{activity_level_mental} reveals that changes in levels of activity are an important signs of various physical and mental health problems like chronic pain and depression. An early and accurate awareness of decreasing activity levels can act as a warning sign to trigger early intervention. Multimodal data fusion over long time periods is a requirement in these types of healthcare applications, and the passive nature of WiFi based CSI sensing could be an important contributor in this area.

\subsection{Taxonomy of WiFi CSI in Healthcare Applications}
From the above we can see that WiFi CSI based methods have some obvious potential to fill the gaps in data left by other sensors. There are however some limitations of this technology and some of advantages and disadvantages are listed below.
\subsubsection{Advantages}
The most obvious advantage of the WiFi CSI based method is that it removes the requirement for wearing sensors. This acts to increase user uptake and captures activities and behaviors in natural rather than artificial conditions. Secondly WiFi based systems avoid many of the the privacy concerns associated with other sensors such as CCTV. Thirdly, within the coverage area of the WiFi signal, the WiFi CSI based method can provide panoramic monitoring independent of light conditions. Finally, recent progress in  signal processing and machine learning has made it possible to extract detailed behavior information from WiFi CSI data. In this paper, we emphasize the frequency component in CSI as it is directly related the motion of a subject of interest. 

\subsubsection{Disadvantages}
Although WiFi CSI based behavior recognition is a promising approach there are some limitations of the technology. Firstly, lots of current WiFi based CSI gesture/activity recognition approaches are carried out in highly controlled environments. In practice changes of geometry may also degrade the recognition performance. Secondly, the WiFi CSI based research mentioned in the previous sections require the WiFi AP and clients to work in data transmission mode in order to obtain the OFDM signal needed for the processing. This condition cannot be guaranteed in many situations as the WiFi AP may only broadcast a beacon signal in 10 short bursts over 1 second if the WiFi AP is in idle status. Even when the WiFi network in being used, there may only be very sparse WiFi transmissions that are insufficient to accurately estimate the required parameters. It is therefore unlikely that one single technology will meet all challenges in residential healthcare. In practice, WiFi CSI based behavior recognition data needs to be integrated and fused with data from other sensors for better activity recognition performance or to fill the gaps in time and location coverage for continuous and seamless monitoring.

\section{WiFi CSI signal Processing and Behavior Recognition Methods} \label{s:dsp_br}

\begin{figure}[hb]
\centering
\includegraphics[width=\linewidth]{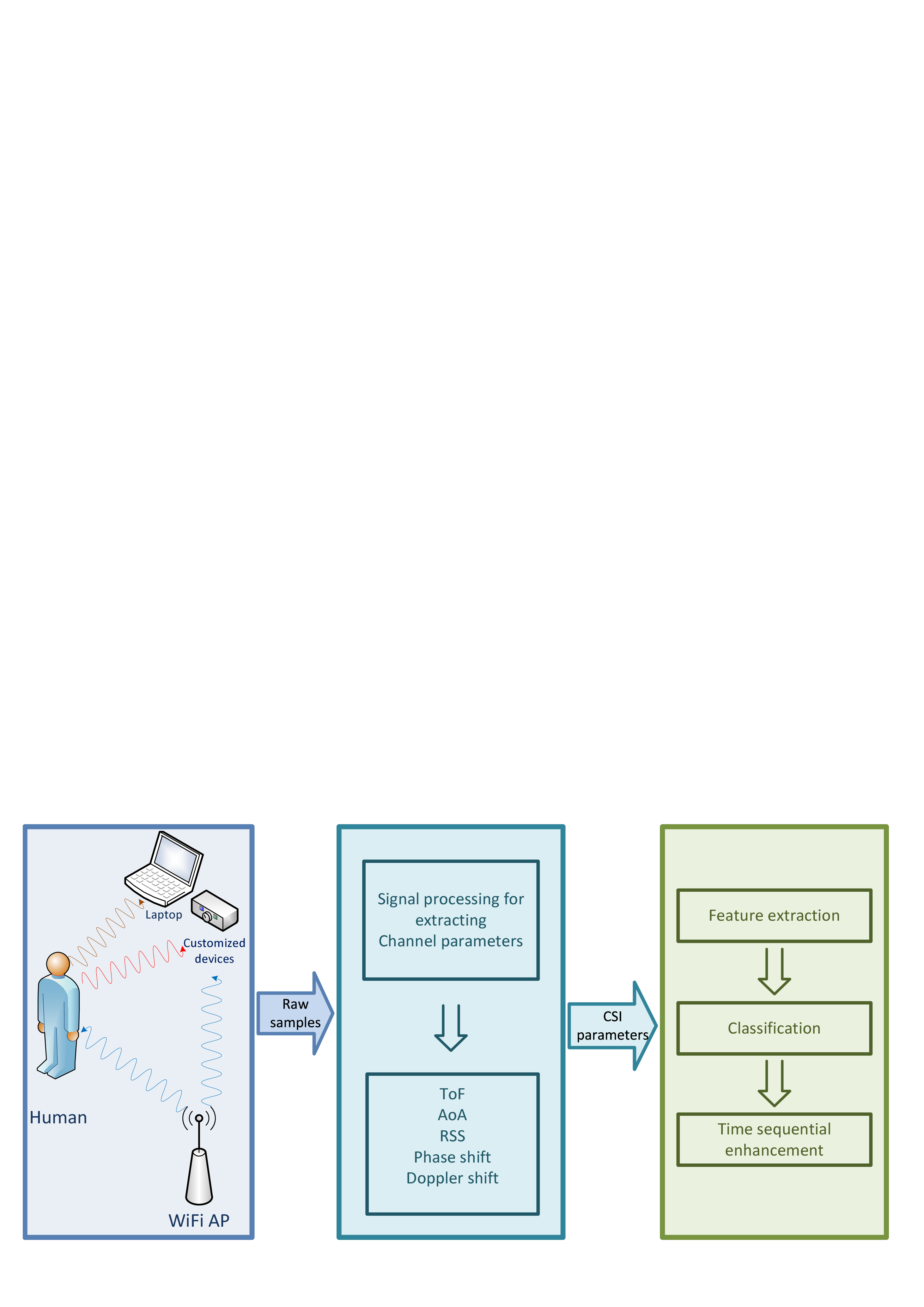}
\caption{The general composition and process of WIFi CSI based human behavior. (a) Signal capturing, (b) Signal processing for CSI extraction, (c) Machine learning to recognize the behavior.}
\label{fig:Top_level}
\end{figure}

Both signal processing and machine learning techniques are used in conjunction with WiFI CSI measurements for behaviour recognition and activity monitoring. In this section we outline various approaches which have been employed to extract WiFi CSI, including those which have been used to identify frequency (Doppler) shifts caused from physical body movements in the vicinity of a WiFi AP. We also discuss methods used to classify CSI/Doppler signatures inherent to a given motion, and the potential improvements afforded by making use of temporal patterns within the frequency data. Figure \ref{fig:Top_level} illustrates the main processing and classification steps in scenarios where people move through WiFi fields. 

\subsection{Extracting the WiFi CSI}
\subsubsection{Commercial Off-the-Shelf Devices}
The majority of commercial WiFi-enabled devices are able to parse WiFi signal data and output information about the state of the channel, the most common being the received signal strength indicator (RSSI). However factors such as the orientation of scatters, multipath and shadowing act as major sources of error in RSS measurements. Techniques such as 'Fingerprinting' therefore require initial characterization procedures within an environment, prior to carrying out any localization tasks. 

More recently, researchers have taken advantage of the Intel 5300NIC to capture WiFi CSI. It uses pilot OFDM symbols in 802.11n signal to estimate the CSI, and reports the channel matrices for 30 subcarrier groups from 3 receiving antennas. Each matrix consists of complex entries with signed 8-bit resolution each for both the real and imaginary parts. Each entry can be written as $|h_{ij}|e^{-j\phi_{ij}}$, where $h_{ij}$ and $\phi_{ij}$ are the channel amplitude and phase property of the $i_{th}$ receiving antenna and $j_{th}$ subcarrier. The measured phase on each receiving antenna and subcarrier provides an opportunity to use well-established signal processing methods, a popular technique being the subspace based joint angle and time estimation technique on the basis of Schmidt Orthogonalization \cite{Schmidt} and \cite{SpotFi}. 

\subsubsection{Dedicated Devices} \label{ss:CSF}
In addition to COTS solutions for extracting WiFi CSI, bespoke systems dedicated to isolating more detailed information relating to only 1 or 2 channel parameters have appeared in the literature. In these cases software defined radio (SDR) systems have typically been used to acquire raw WiFi data in order to apply customized signal processing for extracting CSI. For example, xDtrack \cite{Matrack} applies a subspace search method on raw IQ samples to determine phase variations for high-resolution ToA and AoA estimates. By taking more IQ samples, passive WiFi sensing \cite{Passive_WiFi_TAES} is able to ascertain channel Doppler shifts at very high resolution as well. The signal processing method in \cite{Passive_WiFi_TAES} employs the cross ambiguity function (CAF) to compare the original WiFi transmission with measured reflections to identify small Doppler shifts resulting from moving people. Particularly for Doppler shift and frequency component, there are two main approaches. One approach is to apply STFT or DWT on CSI for extracting frequency component. However, this approach has a limitation on discriminating moving target reflections and stationary reflections. Another approach which is based on passive radar principle applies CAF processing on sampled signals from reference and surveillance channels that contains stationary source signal and moving target reflection respectively. The passive radar approach shows good performance on cancelling the impact from stationary reflection. In Section IV we present a high resolution passive WiFi Doppler radar system and experimental recognition results.

\subsection{Behavior Recognition}
\subsubsection{Temporal Channel State Variations}
\begin{figure}
\centering
\includegraphics[width=\linewidth]{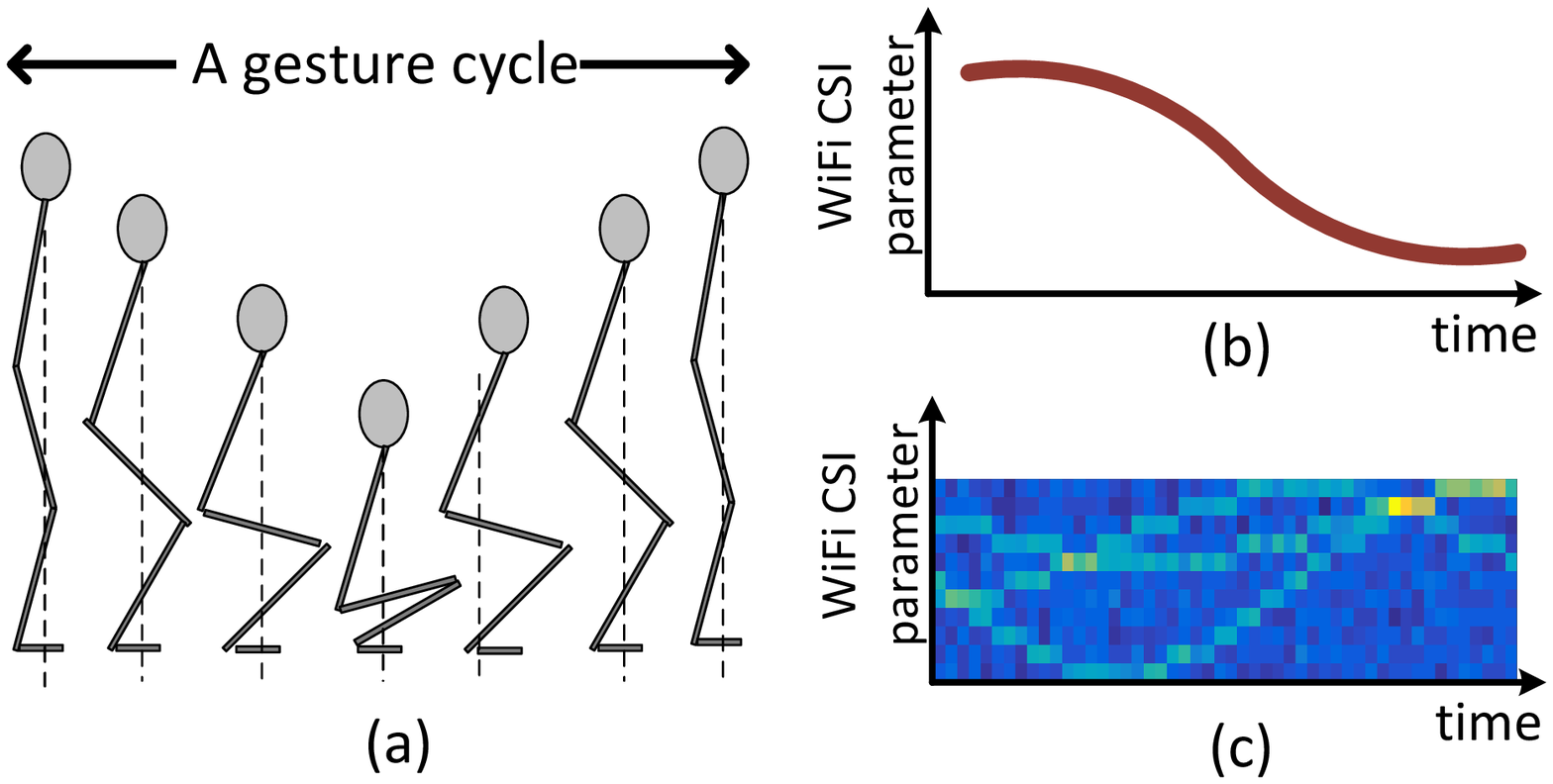}
\caption{The parameter change during an example gesture cycle. (a) Example gesture cycle, (b) Change of recognizable parameter value during one gesture cycle, (c) Change of the unrecognizable parameter value during one gesture cycle.}
\label{fig:example_gesture}
\end{figure}

At any given instant, a measurement of the WiFi channel state provides little information relating to the behaviour of a person in the signal propagation path. However, analyzing the CSI as a function of time offers this possibility: a physical change in a persons position, speed and/or direction will affect propagation paths, Doppler shifts and arrival angles. The effect is a distortion in the channel with a characteristic temporal signature. Even if a person attempts to stand still, the movement of their torso (e.g. swaying, bending etc) or limbs (e.g. lifting a cup etc) will give rise to time-varying commensurate changes in the phase, frequency and amplitude characteristics of the reflected signal. In practice a typical gesture cycle such as that shown in Figure \ref{fig:example_gesture} will be made up of a combination of motions from various parts of the body, leading to a complex temporal signature. We find that particular CSI parameters actually show a clear change in their temporal trace during one gesture cycle when there is one dominant motion and minimal interference and noise - see Figure \ref{fig:example_gesture} (b). However, as illustrated in Figure \ref{fig:example_gesture} (c) when there are multiple contributory motions during a gesture cycle, CSI parameters change in a more complex manner. 

\subsubsection{Feature Extraction}

The performance of a classifier for recognizing activities and behaviours depend highly on appropriately defining and extracting discriminate features within a gesture cycle measurement. Our method focuses on extracting (i) intuitive time domain features; (ii) empirical features determined from characteristics of the time domain signal; and (iii) features derived from matrix analysis methods. The time domain features are always the correlations among time domain signals, with the consideration of time mis-alignment among intra-class samples. 

As can be seen from Figure \ref{fig:example_gesture}(b), some CSI parameters show a recognizable pattern during the gesture cycle. The selection of features to use such as the peak value, span, slope, zero-crossing and inflection points of a measured curve must be based on empirical data. While it is impossible to extract intuitive features in a general case as shown in Figure \ref{fig:example_gesture}(c), matrix analysis can be utilized to analyze structural properties during a gesture cycle, and typical approaches include Principle Component Analysis (PCA) and the Singular Value Decomposition (SVD). 

\subsubsection{Classification of Gestures/Activities}

Many healthcare applications require real-time classification of a gesture or activity, such as falling down, so that an alert can be instantly triggered. A number of classifiers are able to meet this requirement but their applicability also depends on the type of input features chosen. In general, the most popular classification methods are the Naive Bayesian classifier, support vector machine (SVM) classifier and the sparse representation classifier (SRC). The Naive Bayesian classifier is simple and flexible to feature types but only suitable for small number of features. SVM is widely used for different problems because it is applicable to both linear and nonlinear data. In this work we utilize SRC approaches because of its robustness to low signal to noise ratios, and misaligned data. However, SRC is more complex than SVM and other commonly used classifiers.

\section{Case Study} \label{a:case_study}
In Section \ref{s:challenges}, we described the challenges of human behavior recognition in residential healthcare. This included detecting vital signs, potentially fatal events, daily activity recognition and longer term pattern-of-life monitoring. In this section, we present three cases supported with practical experimental data to demonstrate how WiFi CSI based human behavior recognition, especially Doppler and phase variations, fit these challenges in residential healthcare.

\subsection{Case 1: Through-Wall Detection of Vital Signs} \label{sc:respiration}
\begin{figure}[hb]
 \centering
  \includegraphics[width=\linewidth]{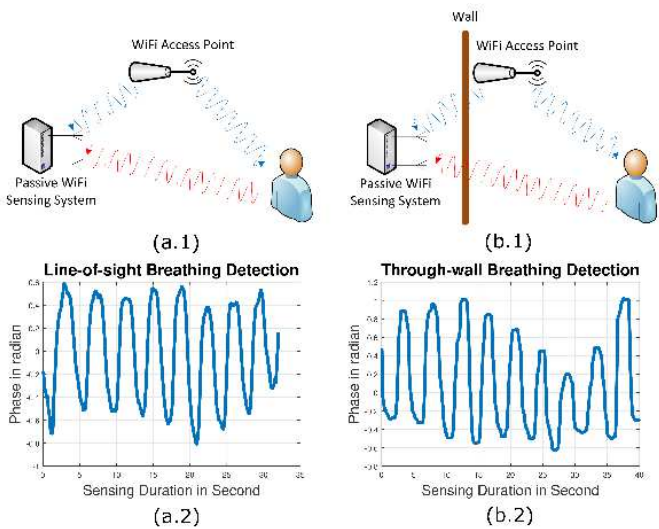}
 \caption{The line-of-sight and through-wall respiration sensing based on WiFi phase measurement. (a.1), Layout of LoS WiFi respiration sensing, (a.2), LoS respiration detected result, (b.1) Layout of through-wall WiFi respiration sensing, (b.2), Through-wall respiration detected result}
 \label{fig:breath}
\end{figure}

As mentioned in Section \ref{s:dsp_br}, useful medical information can be identified during a gesture cycle. For example, doctors may want to know the respiration rate of patients during sleeping. The dominant chest-wall movement caused by respiration has a  0.5 to 2 centimeter displacement. This small movement will generate a 0.25 to 1 radian phase change on the 2.4 GHz WiFi signal, or generate a 0.6 to 2.4 radian phased change on a 5.8 GHz WiFi signal. In this case, we can leverage the home WiFi signal to provide a non-contact respiration monitoring solution. With the experimental setting in \cite{UCL_breath}, detailed phase variation can be extracted from cross-correlated reference and surveillance signals with help from a Hampel filter which removes the outliers caused by phase noise. passive WiFi sensing has been demonstrated to detect human respiration, even through a 33cm brick wall as shown in Figure\ref{fig:breath}. These results demonstrate that we can observe accurate respiration behavior without the use of complex classification algorithms.

\subsection{Case 2: Daily Activity Recognition in Residential Home} \label{sc:activity_recognition}
\begin{figure}[ht]
 \centering
  \includegraphics[width=\linewidth]{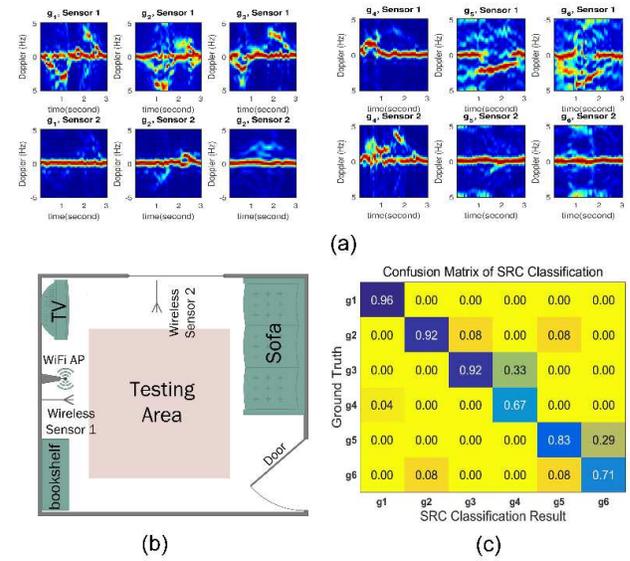}
   \caption{SRC classification of 6 daily behaviors in SPHERE House. (a) Doppler spectrum of gestures captured by two wireless sensors in different angles, (b), SPHERE house experiment layout, (c) SRC recognition result based PCA feature extraction.}
  \label{fig:Doppler_SPHERE}
 \end{figure}

We have applied machine learning methods to recognize a number of behavior patterns in a residential healthcare situation. The experiments were carried out in the SPHERE house \cite{SPHERE_book} (Bristol, UK) using a COTS WiFi AP and SDR based passive WiFi radar. The CAF function \cite{Passive_WiFi_TAES} was applied to extract small Doppler shifts, and the results obtained are summarized below. First, a group of gestures that are deemed important in residential healthcare were selected. The gestures are as follows: \textbf{$g_1$}: \textit{picking up an item from the floor}, \textbf{$g_2$}: \textit{sitting  down on a chair}, \textbf{$g_3$}: \textit{standing up from a sitting position}, \textbf{$g_4$}: \textit{falling down on the floor}, \textbf{$g_5$}: \textit{standing up after a fall}, \textbf{$g_6$}: \textit{get up and out of a bed}. The CAF method introduced in Section \ref{ss:CSF} is applied to a two receiver passive WiFi sensing system to extract accurate and high resolution frequency (Doppler) perturbations cause by human movement. The extracted Doppler traces from the two sensors during each gesture cycle are shown in Figure \ref{fig:Doppler_SPHERE} (a). Taking \textbf{$g_1$} for example, picking up an item from floor often includes leaning over, picking and getting up. We can observe from Sensor 1 in the Doppler-time spectrum in Figure \ref{fig:Doppler_SPHERE} (a) that there is an obvious negative Doppler trace, short duration at zero Doppler and positive Doppler shifts. This observation illustrates a clear correlation between activity and the Doppler-time spectrum. To systematically interpret human activity from WiFi Doppler detection, we apply SRC classification on the PCA features of each Doppler signature. The behavior recognition result is shown in Figure \ref{fig:Doppler_SPHERE} (c). Note that the CAF based frequency estimation can also be applied to both WiFi data bursts and WiFi beacon signals for Doppler information extraction.

Falling is an important activity in healthcare applications \cite{RT_Fall}. Falling events often occur during a relatively short period and exhibit large intra-class variations. Important features are easily missed. The use of SRC in fall detection is advantageous as the sparsity constraints exhibit de-noising properties, which are able to enhance the capture of important local features for classification. Although the recognition rate of the fall motion is still not ideal from Figure\ref{fig:Doppler_SPHERE} (c), it can be improved by considering the correlations among the motions before and after the fall that lead to time sequential modelling.

\subsection{Case 3: Activity Monitoring in a Residential Environment and Sequential Inference}
\subsubsection{Activity Monitoring in a Residential Environment}
\begin{figure}[hb] 
\centering
\includegraphics[width=\linewidth]{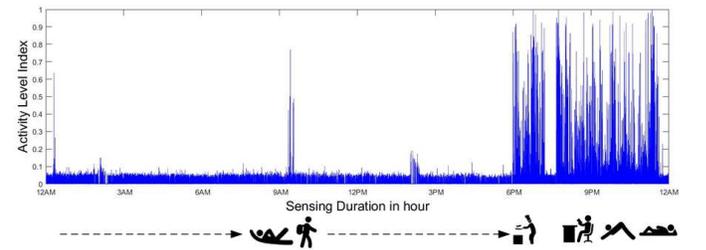}
\caption{An example of 24 hours monitoring based WiFi in a residential house}
\label{fig:24_hours_monitoring}
\end{figure}

As discussed in Section \ref{sc:activity_recognition} and \ref{sc:respiration} the sensing system can detect vital signs by extracting phase information in WiFi CSI and recognize different daily activities by discriminating frequency signatures. These capabilities provide an opportunity to monitor the residential activities in a longer time scale from days to weeks even months and beyond. Figure \ref{fig:24_hours_monitoring} shows 24 hours activity monitoring of a person at home with only COTS home WiFi AP beacons, on the basis of CAF frequency estimation.

The vertical axis in Figure \ref{fig:24_hours_monitoring} shows the activity intensity which correlates to the activity velocity and the relative size of the body. For example walking and the motion of a torso leads to higher intensity values than those produced by arm and head movements. The envelope of the trace is directly related to the activity of the subject. Thus, the intensity data acts as a good candidate to assess the daily lifestyle and activity level. By defining 3 levels according the vertical axis values: sedentary (0.0$\sim$0.3), moderate (0.3$\sim$0.7) and vigorous action (0.7$\sim$1.0). The summary of the person's physical activity intensity is presented as in Table \ref{table:activity_level}. 

\begin{table}[hb]
\centering
\caption{Activity Level Statistics of Daily life}
\label{table:activity_level}
\begin{tabular}{lllll}
Class & \multicolumn{1}{l}{Sedentary} & \multicolumn{1}{l}{\begin{tabular}[c]{@{}l@{}}Moderate \\
Action\end{tabular}} & \multicolumn{1}{l}{\begin{tabular}[c]{@{}l@{}}Vigorous\\ Action\end{tabular}} & \multicolumn{1}{l}{\begin{tabular}[c]{@{}l@{}}Total\\ Action\end{tabular}} \\ \hline
\begin{tabular}[c]{@{}l@{}}Amplitude of\\ activity level\end{tabular} & 0-0.4 & 0.4-0.7 & 0.7-1 & 0-1 \\\hline
\begin{tabular}[c]{@{}l@{}}Time of \\ activity (mins)\end{tabular} & 662.5 (122.5) & 156.5 & 83 & 902 (362) \\ \hline       
\end{tabular}
\end{table}

\subsubsection{Inferring Sequence of Activities} 
Thus far we have discussed instantaneous activity recognition, insofar various classifiers have been applied to the Doppler data, and the predictions were made "batch by batch". Examples of such activities are shown at the bottom of Figure \ref{fig:24_hours_monitoring}. It is accepted \cite{Sequential_Model} that a sequence of activities is correlated in a sense that certain activities are more likely to follow others, and likewise, some are impossible to occur one after another, e.g. a person cannot transition from sitting to running without standing first.     
Such a sequence of activities can be represented by Hidden Markov Model (HMM) (or related models such as Gaussian Random Processes or Conditional Random Field (CRF)). These  sequential methods incorporate future and past observations (here Doppler information) to improve predictions for the current estimate. It is widely reported in other areas where HMM are used (e.g. speech recognition) that such strategies can lead to improved performance compared to techniques that rely only on current observations.   
In the context of residential healthcare, sequential prediction of activities can plausibly help in prediction of future activities or warn of increased risk of a given event.

\section{Conclusions} \label{s:conclusion}
In this article we explain how WiFi CSI can address the challenges of behavior recognition and activity monitoring in residential healthcare. State-of-the-art signal processing techniques make it possible to extract accurate Doppler data, allowing us to characterize activities and behaviours of monitoring huamn subjects. We have presented three case studies to illustrate the capabilities of a passive WiFi Doppler sensing system within a healthcare setting including: monitoring of vital signs, fall detection, and pattern-of-life monitoring. The result confirm that WiFi based CSI sensing technologies show good potential for healthcare applications. We have also identified four key challenges that must be overcome to facilitate a transition of the techniques into real-world assisted living applications. \textbf{(i) WiFi Signal Processing:} Many existing techniques SpotFi \cite{SpotFi} and Matrack \cite{Matrack} rely on high data-rate OFDM WiFi transmissions to extract CSI data. However, only a few methods such as \cite{Passive_WiFi_TAES} can be applied to lower bit-rate or WiFi beacon signals, albeit with performance deterioration. High-resolution algorithms for processing low date-rate signals are therefore critical. \textbf{(ii) Time Alignment:} The classification performance has been shown to depend on accurate time alignment during the gesture cycle. Thus, determining the time points of of start and end of a gesture or behavior is crucial to performance. \textbf{(iii) Multiple Users:} Previous studies such as \cite{RT_Fall}, \cite{Passive_WiFi_TAES}, \cite{WiFiU} and \cite{CARM} have only considered sensing individuals, and extrapolate to multiple users by proposing additional devices. However, additional users significantly add to the sensing complexity in terms of shielding and multipath, sensor deployment, cost etc and is a challenging next step. Finally, \textbf{Sensor Fusion} has shown promising results in WiFi CSI based behavior recognition. It is generally accepted that a variety of sensors will be deployed in future smart homes. Fusing WiFi CSI based recognition data with other sensors like cameras, accelerometers, or electricity, humidity and water meters will provide more accurate, and seamless recognizing and modelling of human behavior.

\bibliographystyle{IEEEtran}

\newpage
Dr Bo Tan is a Lecturer at School of Computing, Electronics and Mathematics, and member of Research Centre for Mobility and Transportation at Coventry University. His research focuses on signal processing for radar and wireless communications systems, and wireless sensing applications in healthcare, security, robotics and indoor positioning. His research in passive WiFi sensing and has led to a series of IEEE conference and journal publications, industrial awards and patent. 

Qingchao Chen received the B.S. degree in Telecommunication Engineering with Management in 2009 from Beijing University of Posts and Telecommunications, Beijing, China. He is currently pursuing his PhD degree in University College London. His research includes MIMO radar system and phased array design, MIMO imaging algorithms, micro-Doppler classification and domain adaptation algorithms.

Karl Woodbridge is Professor of Electronic and Electrical Engineering at University College London. Research interests include multistatic and software-defined radar systems, passive wireless surveillance and micro-Doppler classification. Current research is focused on passive wireless movement detection for Healthcare and Security applications. He is a Fellow of the IET, a Fellow of the UK Institute of Physics, a Senior Member of the IEEE and has published or presented over 200 journal and conference papers.

Dr Kevin Chetty is a Senior Lecturer (Associate Professor) at University College London. His research focuses on RF sensors and signal processing techniques that exploit wireless communications for passive sensing and behaviour classification using micro-Doppler signatures. Other research interests include through-the-wall radar, software defined sensor systems and reconfigurable antennas. He is author of over fifty peer reviewed publications and has been an investigator on grants funded by both government and industry.

Wenda Li received the MEng degree from the Department of Electrical Engineering, University of Bristol, in 2013. Currently, he is a PhD student in the Department of Electrical Engineering, University of Bristol. His research interests include context awareness, system research and location determination system.

Dr Robert Piechocki is a Reader at the University of Bristol. His research interests span all areas of wireless connectivity and sensing, with emphasis on applications to eHealth and transportation. He has published over 120 papers in peer reviewed international journals and conferences and holds 13 patents in these areas. Robert is leading the development of wireless connectivity and sensing for the IRC Sphere project (winner of 2016 World Technology Award).

\ifCLASSOPTIONcaptionsoff
  \newpage
\fi

\end{document}